\documentclass[letterpaper,useAMS,usenatbib, referee]{mn2e}
\usepackage{graphicx}

\newcommand{\be}{\begin{equation}}
\newcommand{\ee}{\end{equation}}
\usepackage{graphicx}
\title[C-Rich Binary Remnants of Extremely Low-Metal AGB Stars]{Carbon Rich Extremely Metal Poor Stars: Signatures of Population-III AGB stars in Binary Systems}
\author[H.B. Lau, R.J. Stancliffe \& C.A. Tout]{Herbert H.B. Lau\thanks{E-mail:
HBL21@ast.cam.ac.uk}, Richard J. Stancliffe  and Christopher A. Tout\\
Institute of Astronomy, The Observatories, Madingley Road, Cambridge CB3 0HA}
\begin{document}
\bibliographystyle{mn2e}
\date{Accepted 2007 March    00. Received 0000 December 00; in original form 0000 October 00}

\pagerange{\pageref{firstpage}--\pageref{lastpage}} \pubyear{0000}

\maketitle

\label{firstpage}

\begin{abstract}
We use the Cambridge stellar evolution code STARS to model the evolution and nucleosynthesis of zero-metallicity intermediate-mass stars. We investigate the effect of duplicity on the nucleosynthesis output of these systems and the potential abundances of the secondaries. The surfaces of zero-metallicity stars are enriched in CNO elements after second dredge up. During binary interaction, such as Roche lobe overflow or wind accretion, metals can be released from these stars and the secondaries enriched in CNO isotopes. We investigate the formation of the two most metal poor stars known, HE 0107-5240 and HE 1327-2326. The observed carbon and nitrogen abundances of HE~0107-5240 can be reproduced by accretion of material from the companion-enhanced wind of a $7\,M_\odot$ star after second dredge-up, though oxygen and sodium are underproduced. We speculate that HE 1327-2326, which is richer in nitrogen and strontium, may similarly be formed by wind accretion in a later AGB phase after third dredge-up.
\end{abstract}


\section{Introduction}

Mass loss from primordial stars is very important because any metals released from them enrich the interstellar medium(ISM), increasing the metallicity of subsequent generations of stars. However mass loss from single stars, though not well understood, is thought to be small compared to solar metallicity stars. To be efficient radiation-driven stellar winds require metal lines \citep*{Marigo}. Thus the winds of zero-and very low-metallicity stars may be very weak, owing to the absence of metals at the surface. However  we do not know how the mass loss depends on metallicity during the asymptotic giant branch (AGB) phase, so it is very uncertain how much this is affected by the low metallicity. On the other hand, mass loss from stars during binary interaction is more well-defined and understood. If the mass transfer is not conservative, the surroundings are enriched in metals. If some of the material lost by the primary is deposited on to the secondary then its surface abundance becomes CNO rich but otherwise metal-poor. This could be one way to form the observed carbon-enhanced metal-poor stars and, at the same time, make a significant contribution to the contamination of the ISM. Observed carbon-enhanced metal-poor stars have a large overabundance of carbon, $ [\rm{C}/\rm{Fe}]> +1.0 $, and they constitute 20\% of stars with $[\rm{Fe}/\rm{H}] \leqslant -2.0$ \citep{Beers}. This suggests a significant amount of carbon was produced in the early Universe. Binary interaction is one possible way to create carbon-rich stars. Alternatively, it is also possible that some carbon-rich stars formed from material enriched by type-II supernovae \citep{Umeda}.

\subsection{Possible formation of binary intermediate-mass primordial stars}

The primordial, first generation of stars, commonly referred to as Population-III stars or zero-metallicity stars, should have the composition of the ISM just after Big Bang nucleosynthesis and hence should be devoid of metals. It has been a popular belief that, in the absence of heavy elements and dust grains, cooling mechanisms are inefficient and favour the formation of massive or very massive stars \citep*{Bromm} because the Jeans mass would be large. However it has also been shown \citep*{Palla} that even a small fraction of molecular hydrogen can provide a significant contribution to the cooling via rotational and vibrational transitions. The resulting Jeans mass of a pure H and He cloud is then relatively small and may even fall below $0.1\,M_\odot$. Owing to this complexity and our lack of understanding of star formation, the initial mass function (IMF) of zero-metallicity stars remains uncertain. \shortcite{Nakamura} suggested a bimodal IMF with peaks around $1\,M_\odot$ and $100\,M_\odot$. Hence, intermediate-mass zero-metallicity stars should be investigated.

\shortcite*{Saigo} explored the formation of Population-III binaries, by simulating the collapse of a rotating cylinder with a three-dimensional, high-resolution nested grid. They found that low angular momentum is key to the formation of binaries. If the initial angular momentum is small enough the runaway collapse of the primordial gas cloud stops at some point and forms a rotationally supported disc instead of continuing to collapse into a single core. After accreting the surrounding cloud, the disc can eventually undergo a ring instability and fragment into a binary. However, if the initial angular momentum is relatively large, a bar-type instability arises and the cloud collapses to a single star through rapid angular momentum transfer. Their results showed that a significant fraction of Population-III stars is expected to form in binary systems, regardless of their mass. Hence the effect of binaries on the evolution and nucleosynthesis of primordial stars can be very important.

\section{The STARS code}
We use the Cambridge stellar evolution code {\sc STARS} to calculate the evolution of primordial intermediate-mass stars. This program was originally written by \shortcite{Eggleton} and has been updated by many authors. The last major update was undertaken by \shortcite{Pols}. It solves the equations of stellar structure, nuclear burning and mixing simultaneously, unlike most codes which treat mixing in a separate step.  Another unique feature of the STARS code is the use of a self-adaptive non-Langrangian, non-Eulerian mesh.  \shortcite{STARSopacity} updated the opacity tables to use the latest OPAL opacities \citep{Iglesias} and these new tables also account for the change in opacity with variations in the carbon and oxygen abundances. We use their zero-metallicity opacities in this work (Eldridge, private communication; \citealt{Ferguson}). 
The nucleosynthesis subroutines of the STARS code were recently resurrected by \shortcite{STARSNucleo}. These make use of the output from the structure code to determine the mixing and nucleosynthesis. In addition to the seven main isotopes that are followed in the structure code (${}^{1}\rm{H},{}^{3}\rm{He},{}^{4}\rm{He}, {}^{12}\rm{C}, {}^{14}\rm{N},{}^{16}\rm{O},{}^{20}\rm{Ne}$) the evolution of 37 other isotopes, including all stable and a few unstable isotopes from ${}^{2}\rm{D}$ to ${}^{34}\rm{S}$ and the major iron group elements, are currently modelled by the nucleosynthesis code.

\section{Evolution of single stars}
We have modelled $3 - 7\,M_\odot$ stars from the pre-main sequence to the beginning of thermally-pulsing AGB using 999 meshpoints without any convective overshooting. The mixing-length parameter \citep{Bohm} $\alpha$ is 1.925 based on calibration to a solar model. The helium mass fraction is chosen to be 0.25 to reflect the prediction of primordial helium abundance from the observed deuterium abundance, baryon density and a spectroscopic sample of extragalactic H-II regions \citep{Fukugita}. The evolution of our models is in qualitative agreement with the previously published work of \shortcite{Chieffi} and \shortcite*{Siess}, even though they used a different value of the initial helium abundance.

We now briefly describe the overall characteristics of the evolution of intermediate mass zero-metallicity stars. The evolution of zero-metallicity stars differs significantly from that of population-I and -II stars because of the absence of carbon, nitrogen and oxygen. Hydrogen cannot burn through the CNO cycle so it burns only by the proton-proton chain. This is less temperature dependent so zero-metallicity stars are considerably hotter in the nuclear-burning regions and throughout most of the star. Their main-sequence lifetimes are much shorter and the burning is much less centralised. Furthermore the temperature at the core becomes hot enough that carbon is slowly produced from helium through the triple-$\alpha$ reaction. At this time there are protons at high temperature so the products of helium burning end up mainly as nitrogen in CNO equilibrium. When the central carbon mass fraction reaches about $10^{-12}$ hydrogen burning switches to the CNO cycle. For stars more massive than $1.8\,M_\odot$, the core is hot enough for helium to burn non-degenerately when central hydrogen is exhausted. This mass, which is the dividing line between low and intermediate-mass stars, is much lower than for solar-metallicity stars because zero-metallicity stars are much hotter and hence have less degenerate cores. The intermediate-mass stars remain on the blue side of the Hertzsprung-Russell (H-R) diagram throughout central helium burning and so the red-giant phase and first dredge up are absent. The surface abundances are unchanged during the core He-burning phase. The upper mass limit for AGB stars at $Z=0$ is between $7\,M_\odot$ to $8\,M_\odot$ because our $8\,M_\odot$ model starts carbon ignition at the core before thermal pulses.

When helium is depleted in the core the stars undergo a large expansion and move rapidly to the red side of the H-R diagram. Hydrogen and helium burning are active in two separate shells. The energy generated by the helium shell causes expansion and cooling of the outermost layers and the convective envelope deepens. This is second dredge-up, although it is actually the first episode of dredge-up that occurs in zero-metallicity stars above $1.8\,M_\odot$.

\begin{figure}
\includegraphics[width=\textwidth]{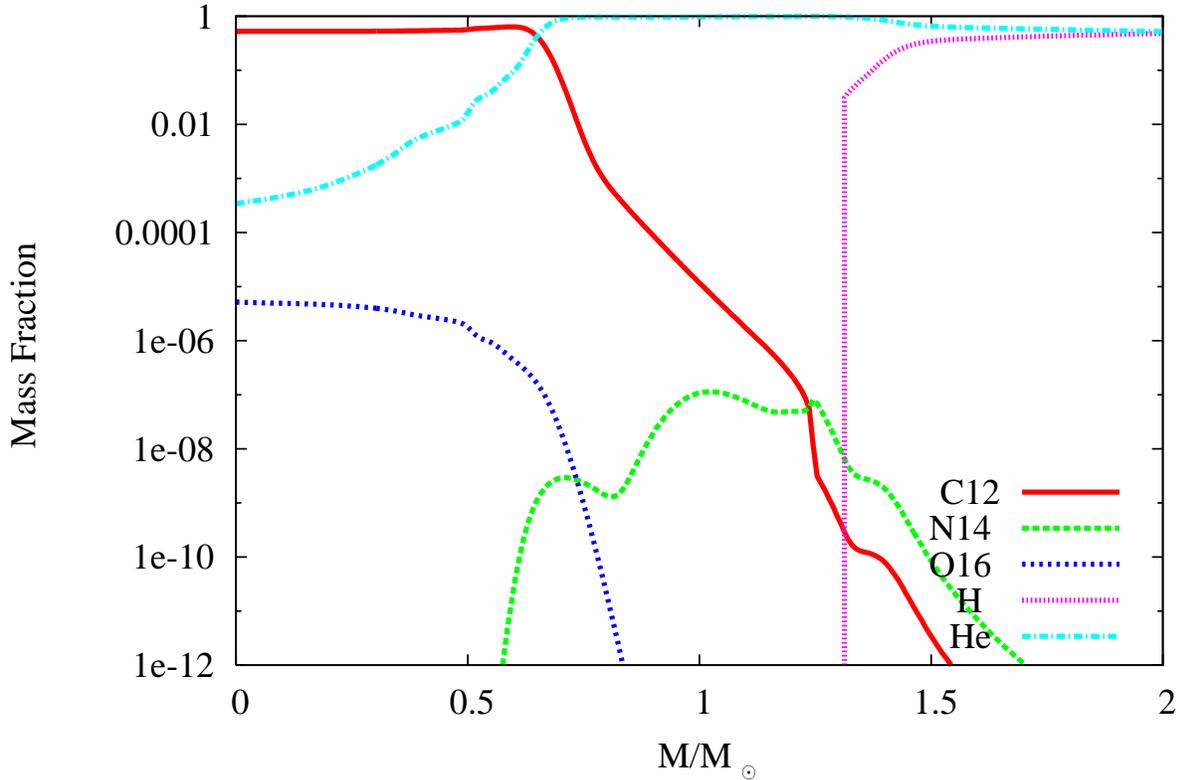}
\caption{Abundances by mass fraction profile of the inner part of a $7\,M_\odot$ model near the end of core helium burning.}     
\label{fig:7mass}
\end{figure}

Enough carbon is produced that hydrogen is mainly burned through the CNO cycle in the H-burning shell and nitrogen is produced therein. Abundance profiles for a $7\,M_\odot$ star are shown in Figure \ref{fig:7mass}. When the convective envelope deepens both carbon and nitrogen are dredged to the surface. 
Lower-mass stars dredge up less of the CNO isotopes because their convective envelopes do not penetrate so deeply and the CNO abundance in the H-burning region is lower because the star is cooler. The surface $\rm{C}/\rm{N}$ ratio (shown in Table 1) is in CNO equilibrium at the temperature of the burning shell. Around 1/100 for lower-mass stars, it increases rapidly from  $5\,M_\odot$ to $7\,M_\odot$. This is because second dredge-up reaches down to the helium-burning shell for higher-mass stars. In addition, the ratio of ${}^{12}\rm{C}/{}^{13}\rm{C}$ increases from 3.5 to 1190. This requires an additional source of ${}^{12}\rm{C}$ that is not present in the CNO H-burning shell but is also explained by this deeper dredge up. Because of the dredging up of materials from the He-burning shell, the carbon abundance increases from $4.07 \times 10^{-11}$ to  $2.64 \times 10^{-6}$ for stars from $5\,M_\odot$ to $7\,M_\odot$. Oxygen follows a similar trend, increasing from $3.56 \times 10^{-11}$ to $3.85 \times 10^{-9}$. Nitrogen is already significantly enhanced at  $2.56 \times  10^{-9}$ for a $5\,M_\odot$ star and only increases a little more to $2.28 \times  10^{-8}$ in the $7\,M_\odot$ model.

\begin{table}
\begin{center}
\begin{tabular}[h] {c c c c c c c c c}
\hline
Mass/$M_\odot$   & ${}^{4}\rm{He}$ &  ${}^{12}\rm{C}$ &  ${}^{13}\rm{C}$ & ${}^{14}\rm{N}$& ${}^{16}\rm{O}$ & ${}^{12}\rm{C}/{}^{13}\rm{C}$ & ${}^{12}\rm{C}/{}^{14}\rm{N}$ & ${}^{12}\rm{C}/{}^{16}\rm{O}$\\
\hline
3.0 & $0.352$& $2.51 \times 10^{-18}$ &$7.18 \times 10^{-19}$     & $3.08 \times  10^{-16}$ & $7.80 \times 10^{-18}$ & 3.50 & $8.15 \times 10^{-3}$ & $3.21 \times 10^{-1}$\\
4.0 & $0.370$& $2.63 \times 10^{-14}$ &$7.46 \times 10^{-15}$     & $2.77 \times  10^{-12}$ & $6.05 \times 10^{-14}$ & 3.53 & $9.49 \times 10^{-3}$ & $4.34 \times 10^{-1}$\\
5.0 & $0.380$& $4.07 \times 10^{-11}$ &$ 1.16 \times 10^{-11}$    & $2.56 \times  10^{-9}$  & $3.56 \times 10^{-11}$ & 3.51 & $1.59 \times 10^{-2}$ & $1.14 $\\ 
6.0 & $0.382$& $9.38 \times 10^{-8}$  & $1.20 \times 10^{-9}$    & $1.73 \times  10^{-8}$ & $1.27 \times 10^{-10}$ & 78.2 & $5.42$ & $7.39 \times 10^{2}$\\
7.0 & $0.382$& $2.64 \times 10^{-6}$  & $2.22 \times 10^{-9}$    & $2.28 \times  10^{-8}$  & $3.85 \times 10^{-9}$ &1190 & $1.16 \times 10^{2}$  & $6.86 \times 10^{2}$\\
\hline
\end{tabular}
\caption{Surface abundances and abundance ratios by mass fraction at the end of the early AGB before thermal pulses.}
\end{center}
\end{table}

\section{Mass loss from AGB stars}
It is generally believed that mass loss from zero-metallity stars is very slow because radiation driven winds are much weaker \citep{Marigo}. However, the mass-loss mechanism during AGB evolution is not well understood. It has been suggested that the increase of the mass-loss rate during the AGB is caused by the onset of radial pulsations (\citealt{VWML, Blocker}) or magnetic fields \citep{Blackman}. Whether the mass loss would then be dependent on metallicity is not known. In addition the lower opacity of a metal-deficient atmosphere generally leads to more compact stars so the mass-loss rate is expected to be lower in any case. However, after dredge up, the surfaces of the stars are also enriched in carbon and rates may increase again. Reimers'  mass-loss rate  \citep{Reimersformula},
 
\begin{equation}
\dot{M}= 4.0\times10^{-13} \eta \frac {(L/L_\odot)(R/R_\odot)} {(M/M_\odot)} M_\odot \,\rm{{yr}}^{-1},
\end{equation}
where $L$ is the luminosity of the star, $M$ its mass and $R$ its radius, reproduces the early AGB mass loss of higher metallicity stars reasonably well when $0.4<\eta<3$ \citep{Straniero}. We use $\eta$ of both 1 and 4 (which almost certainly overestimates the mass loss from $Z=0$ stars) for $3\,M_\odot$, $5\,M_\odot$ and $7\,M_\odot$ models to simulate the mass loss during the  early AGB phase. We define the yield of isotope $i$ by
\begin{equation}
Y_{i}=\int_{0}^{T}\dot{M}(t)X_{i}(t) \,{dt},
\end{equation}
where $ \dot{M}(t)$ is the mass-loss rate, $T$ is the lifetime of the star and $X_{i}(t)$ is the surface abundance at time $t$ of isotope $i$.

As can be seen in Tables 2 and 3,  the yield is directly proportional to $\eta$ because changing the value of $\eta$ has little effect on the stars' evolution because their mass-loss rates remain low. Our $3\,M_\odot$ star releases a negligible quantity (of the order of $10^{-15}$) of metals because the dredge-up is not deep enough to bring many to the surface. Our $7\,M_\odot$ star has significantly higher yields of CNO elements, particularly carbon ($4.82 \times 10^{-8}\,M_\odot$ for $\eta=1$) and to a lesser extent oxygen ($6.77 \times 10^{-11}\,M_\odot$). This is because the convective envelope reaches very close to the helium burning shell to dredge up carbon and this is only possible if the star is more massive than about $5\,M_\odot$. The nitrogen yield for a $5\,M_\odot$ star is $6.95 \times 10^{-11}\,M_\odot$ compared with $3.61 \times 10^{-10}\,M_\odot$ for $7\,M_\odot$. A little carbon is also present outside the helium-burning shell because it was produced at the end of main sequence by the triple-$\alpha$ reaction when the core was hot and extended. Most of the carbon produced then was immediately converted to nitrogen by the CNO cycle. So in order to dredge up nitrogen the convective envelope needs only to reach the region in which the CN cycle was active. This first happens at around $5\,M_\odot$. Given the uncertainty in the initial mass function for zero-metallicity stars, we cannot make an accurate prediction of the relative contribution of different mass ranges. However, because the contribution of carbon from a  $7\,M_\odot$ star is more than 1,000 times that of a $5\,M_\odot$ star, we can see that the higher-mass AGB stars can dominate the carbon contribution for a normal IMF such as that of \shortcite{Salpeter}. Similarly, the oxgyen yield is dominated by the $7\,M_\odot$ stars because it is nearly 100 times greater than that of a $5\,M_\odot$ star. Unless the ratio of $5\,M_\odot$ to $7\,M_\odot$ stars is a few times higher than Salpeter's ratio the nitrogen contribution from $5\,M_\odot$ stars is negligible compared to that of $7\,M_\odot$ stars.

\begin{table}
\begin{center}
\begin{tabular}[t] {c c c c c c}
 \hline
Initial Mass/$M_\odot$  & Mass Lost/$M_\odot$ & $Y_{{}^{12}\rm{C}}$/$M_\odot$ &  $Y_{{}^{14}\rm{N}}$/$M_\odot$& $Y_{{}^{16}\rm{O}}$/$M_\odot$\\
\hline
3.0 & 0.077 &    $1.71 \times 10^{-17}$ & $1.94 \times 10^{-15}$ & $5.05 \times 10^{-17}$\\
5.0 & 0.052 &    $1.22 \times 10^{-12}$ & $6.95 \times 10^{-11}$ & $9.07 \times 10^{-13}$\\
7.0 & 0.034 &    $4.82 \times 10^{-8}$  & $3.61 \times 10^{-10}$  & $6.77 \times 10^{-11}$\\
\hline
\end{tabular}
\caption{Mass loss and yield of CNO before thermal pulses with $\eta =1$ for single star.}
\end{center}
\end{table}
\begin{table}
\begin{center}
\begin{tabular}[t] {c c c c c c}
\hline
Initial Mass/$M_\odot$  & Mass Lost/$M_\odot$ & $Y_{{}^{12}\rm{C}}$/$M_\odot$ &  $Y_{{}^{14}\rm{N}}$/$M_\odot$& $Y_{{}^{16}\rm{O}}$/$M_\odot$\\
\hline
3.0 &       0.32 &$8.01 \times 10^{-17}$& $9.06 \times 10^{-15}$ & $2.36 \times10^{-16}$\\
5.0 &       0.22 &$4.79\times 10^{-12}$& $2.74 \times 10^{-10}$  & $3.63 \times 10^{-12}$\\
7.0 &       0.14 &$1.94 \times 10^{-7}$ & $1.47 \times 10^{-9}$     & $2.70 \times10^{-10}$\\
\hline
\end{tabular}
\caption{Mass loss and yield of CNO before thermal pulses with $\eta =4$ for single star.}
\end{center}
\end{table}

\section{Binary stars}
We expect mass transfer by Roche lobe overflow in low-metallicity binary stars to proceed in a similar way to higher metallicity systems. If the mass transfer is non-conservative the surroundings are also enriched in metals but the degree to which mass transfer is non-conservative is not known. If the system is a very close binary, mass transfer begins on the main sequence. Otherwise, because the radius of a zero-metallicity intermediate-mass star only increases substantially after it leaves the main sequence, it is more likely for the mass transfer to begin in the early AGB phase than before. On the main sequence the star's radius increases by a factor of two while on the early AGB it increases by a factor of one hundred. In this case, when the primary fills its Roche lobe, any mass lost contains a significant amount of metals because dredge up has already occured. In cases when mass transfer begins before second dredge up, such as the  Hertzsprung Gap, CNO isotopes can still be transferred because the CNO-rich H-burning shell is eventually exposed at the surface.

We have made several binary models with primary masses ranging from $5$ to $7\,M_\odot$ and an initial mass ratio of two-thirds, with different periods (see table 4). These models are evolved without any tidal interaction. The stellar evolution of the secondary is also not modelled because a lower mass secondary would not have evolved when mass transfer starts. The secondary star is thus treated as a point mass. The variation in the period of the system is modelled by considering the change in angular momentum during binary interaction.  Mass transfer begins when the primary is in the Hertzsprung Gap for systems shown in the table. Longer periods (not listed) could result in a common envelope because mass transfer starts when the primary's envelope is deeply convective. If tides were included, the periods would decrease because orbital angular momentum would go to spin up the expanding primary. We do not explicitly model tidal effects so our models really correspond to slightly larger initial periods. In the case of a $7\,M_\odot$ primary star the envelope can be completely lost to leave a $1\,M_\odot$ CO white dwarf. Even when $6\,M_\odot$ is transferred to the companion, we are only interested in the mass released from the primary, and so treat the companion as a unevolved point mass. This does not have any significant effect on the result. Provided that the mass transfer rate remains low, the radius increase due to accretion by the secondary would not severely disrupt the binary system. Significant masses of metals are transferred from the primary, up to  $4.00\times10^{-5}\,M_\odot$, $5.23 \times 10^{-7}\,M_\odot$ and $1.93 \times 10^{-7}\,M_\odot$ of C, N and O respectively. Further increases in the orbital period would only have a small effect on the compositions of the material lost because mass transfer starts after the metals have already been dredged up. As the binary period increases the remnant mass slightly increases but this has only a very small effect on the total amount transferred from the primary. 

The mass of metals transferred from the primary star is more than one hundred times higher than the mass of metals released during typical single-star mass loss at the same evolutionary point. In particular the carbon transferred from the primary envelope is nearly 1,000 times greater. This is because the whole envelope of the primary is lost during binary interaction while a very small fraction is lost from a single star by the AGB wind. We have computed a single $7\,M_\odot$ star model through all its 600 thermal pulses or so and found that these are too weak to lead to any third dredge up. Carbon is converted to nitrogen because of hot bottom burning. Later pulses are weaker and they cease altogether when the core mass reaches $1.1\,M_\odot$. The stars then enter a non-eventful evolutionary phase while the hydrogen and helium burning shells grow outward. Eventually, carbon ignites degenerately at the centre when the core mass is $1.36\,M_\odot$. The subsequent thermonuclear runaway (similar to a type Ia supernova) would release sufficient energy to blow the whole star apart. At this point, the surface carbon and nitrogen abundances are $5.5 \times 10^{-7}$ and  $2.7 \times 10^{-6}$. The composition of the lost material would be very different to the binary case discussed previously, with up to  $1.6 \times10^{-5}\,M_\odot$ of nitrogen being ejected while the yield of carbon would be of the order of  $3.3 \times 10^{-6}\,M_\odot$ depending on mass-loss rate. This is over thirty times more nitrogen. Despite the envelope, we would expect the explosion mechanism of this star to be very similar to a Type Ia supernova, so the carbon and oxygen ejected from the supernova later on exceeds the yield from the envelope (Lau, Stancliffe \& Tout, in prep).

We also model a $7\,M_\odot$ primary star with a $4\,$d period with different mass transfer efficiencies, 0.25, 0.5 and 0.75, where the mass-transfer efficiency is the ratio of the mass transferred to the seondary to the mass lost from the primary. A mass transfer efficiency of 0.75 means three-quarters of the mass lost from the primary is accreted by the secondary and the remaining quarter is lost from the system. We calculate the masses of metals released from the binary system and list them in table 5. The increase in the metals released is already significant even when the mass-transfer efficiency is 0.75.  The carbon, nitrogen and oxygen yields are 100 to 1,000 times more than the single star yield of a similar age. 

For the case of mass-transfer efficiency of 0.25, mass transfer proceeds very rapidly because of the greater mass ratio throughout owing to the secondary accreting less material. Because mass is transferred from the more massive primary to the less massive secondary, the Roche lobe radius shrinks, while the radius of the primary increases as mass is lost from its convective envelope. This leads to positive feedback and further increases the mass-loss rate. In practice, this probably leads to common envelope evolution. The common envelope surrounding both stars is likely to be lost from the system, so we assume the remaining primary's envelope is released from the binary soon after the common envelope forms. In this case, the total amount of CNO released is even higher because most of the primary's metal-rich envelope is released from the system. In table 5, we assume all the envelope of the primary is lost from the system at the onset of the common envelope evolution.

\begin{table}
\begin{center}
\begin{tabular}[t] {c c c c c c}
\hline
Primary Mass/$M_\odot$ &  Period/d & Final Mass/$M_\odot$ & $Y'_{{}^{12}\rm{C}}$ /$M_\odot$ &  $Y'_{{}^{14}\rm{N}}$/$M_\odot$& $Y'_{{}^{16}\rm{O}}$/$M_\odot$\\
\hline
7.0 & 1.0 & 1.00366 &   $2.97\times10^{-5}$ & $4.11 \times 10^{-7}$ & $1.52 \times10^{-7}$\\
7.0 & 4.0 & 1.03236 &   $4.00\times10^{-5}$ & $5.23 \times 10^{-7}$ & $1.93 \times10^{-7}$\\
6.0 & 1.0 & 0.95516 &   $8.59\times 10^{-7}$ & $2.67 \times 10^{-7}$ & $2.42 \times 10^{-9}$\\
6.0 & 3.0 & 0.97650 &   $1.01\times 10^{-6}$ & $2.68 \times 10^{-7}$ & $2.92 \times 10^{-9}$\\
5.0 & 1.0 & 0.91337 &  $6.47 \times 10^{-11}$ &$5.20 \times 10^{-9}$ & $9.25 \times 10^{-11}$\\ 
5.0 & 3.0 & 0.93518&  $1.51\times 10^{-10}$ & $1.08\times10^{-8}$ & $1.68 \times10^{-10}$\\
5.0 & 10.0& 0.9406 & $9.27 \times 10^{-11}$ & $6.64 \times 10^{-9}$ & $1.04 \times 10^{-10}$\\
\hline
\end{tabular}
\caption{The masses of metals transferred from the primary $Y'$in various binary interactions. Some is accreted by the companion and rest lost to the ISM}
\end{center}
\end{table}

\begin{table}
\begin{center}
\begin{tabular}[t] {c c c c c c}
\hline
Mass transfer efficiency & $Y_{{}^{12}\rm{C}}$/$M_\odot$ &  $Y_{{}^{14}\rm{N}}$/$M_\odot$& $Y_{{}^{16}\rm{O}}$/$M_\odot$\\
\hline
0.75 & $5.63 \times 10^{-6}$ & $1.60 \times 10^{-7}$ & $2.10 \times 10^{-8}$\\
0.50 & $1.77 \times 10^{-5}$ & $2.44 \times 10^{-7}$ & $8.15 \times 10^{-8}$\\
0.25 & $3.30 \times 10^{-5}$ & $4.55 \times 10^{-7}$ & $1.14 \times 10^{-7}$\\

\hline
\end{tabular}
\caption{The masses of metals released from a binary system with initial primary mass of $7,M_\odot$, $q=2/3$ and period of 4 days.}
\end{center}
\end{table}

\section{Carbon enhanced metal poor stars}
There have been two discoveries of carbon-enhanced hyper-metal-poor stars ($-6.0 \leqslant [\rm{Fe}/\rm{H}] \leqslant -5.0$), HE 0107-5240 (\citealt{Christliebn}; 2004) and HE 1317-2326 (\citealt{Frebel}; \citealt{Aoki}). They are the lowest-metallicity stars observed so far. The important elemental and isotopic abundances are summarized in Table 6.
\begin{table}
\begin{center}
\begin{tabular}[t] {c c c c}
\hline
Abundances & HE 0107-5240 & HE 1317-2326 & $7\,M_\odot$ early AGB models \\
\hline
[$\rm{Fe}/\rm{H}$] & $-5.3$  &  $-5.45$ &  $-5.26$              \\

[$\rm{C}/\rm{Fe}$] & $3.7$   &  $3.90-4.26$ &   $2.73$\\
$\rm{C}/\rm{N}$    & $40-150$& $3$ &    $239$ \\

[$\rm{O}/\rm{Fe}$] & $2.3$   & $<4.0$  &$0.1$ \\ 
${}^{12}\rm{C}/{}^{13}\rm{C}$ & $ >50$  & $>5$   & $14200$  \\

[$\rm{Sr}/\rm{Fe}]$  & $<-0.52$     &$0.91-1.24$ & Sr not enhanced \\
\hline
\end{tabular}

\caption{1D atmospheric analysis abundances and surface abundances of observed CeHMP stars and our $7\,M_\odot$ star with $Z=10^{-7}$ after second dredge-up.}
\end{center}
\end{table}

HE 0107-5240 is not significantly enriched in strontium while the other has an unexpectedly high strontium abundance. Carbon is very much enhanced relative to the solar $\rm{C}/\rm{Fe}$ ratio, by a factor of several thousand. Nitrogen and oxygen are also strongly enhanced, being a factor of more than 100 above the solar $\rm{N}/\rm{Fe}$ and $\rm{O}/\rm{Fe}$ ratios. The  $\rm{C}/\rm{N}$ abundance ratio  of 40 to 150 for HE 0107-5240 is similar to the surface abundances of the early AGB phases of our $6-7\,M_\odot$ stars with $Z=10^{-7}$, with the initial isotopic ratios taken as solar

Because there is already iron enrichment, the progenitors of these stars have not evolved directly from $Z=0$ stars. We therefore carried out a new set of calculations at $Z=10^{-7}$. For these intermediate mass stars this does not introduce enough CNO isotopes to qualitatively alter their evolution from those of $Z=0$ but we model a new set at $Z=10^{-7}$ for consistency. The observed ${}^{12}\rm{C}/{}^{13}\rm{C}$ ratio, which is greater than 50, is not inconsistent with the models. We do not include strontium in our code but we can deduce that strontium is not enhanced in the early $7\,M_\odot$ AGB stars because of the absence of a neutron source. One possible formation scenario for HE~0107-5240 is that it is the secondary star of a wide binary system, of which the original primary star has already evolved into an unseen white dwarf. This is very similar to the best explanation for Ba stars of solar metallicity \citep{Karakas}. Mass was transferred to the secondary through wind accretion during the early AGB phase of the primary. \shortcite{Suda} proposed an alternative scenerio in which the primary star has an initial mass between 1.2 and $3\,M_\odot$. The major difference is that, in their lower mass models, the controversial helium-flash-driven convective mixing is responsible for bringing metals to the surface, while the well-established second dredge up alone is enough to increase surface carbon and nitrogen in our $7\,M_\odot$ models. 

In order to explain the non-detection of radial velocity variation,  the binary has to be wide enough to avoid Roche lobe overflow and common envelope evolution. Otherwise, the final period would be low enough for the companion to be detectable. The minimum initial period of the binary system must have been at least $600\,$d. Because of the loss of angular momentum in the wind, such a period would increase to at least $30\,$yr at the present day. Moreover in our model the current radial velocity of the secondary would be about $6.5\,\rm{km}\,s^{-1}$. This theoretical value with the $30\,$yr period can explain the non-detection of variation in radial velocity of HE 0107-5240 \citep*{Bessell}. The maximum velocity variation would be $4.8\,\rm{km}\,s^{-1}$ after 4 years of observation with radial velocity variation of up to $13\,\rm{km}\,s^{-1}$ in 15 yr. Considering the orientation of the orbit, we might expect a velocity variation of about $1\,\rm{km}\,s^{-1}$ after 4 years and $5\,\rm{km}\,s^{-1}$ in 15 yr.

For systems that are too wide for Roche lobe overflow, \shortcite{Tout} suggested that mass loss may be tidally enhanced by the presence of a moderately close companion. Their enhanced mass loss formula is

\begin{equation}
\dot{M}=\dot{M_{R}}\left\{ 1+B ~\textrm{max}\left( \frac{1}{2},\frac{R}{R_{\rm L}}\right)^6\right\},
\end{equation}
where $\dot{M_{R}}$ is the original Reimers mass-loss rate for a single star, $R_{\rm L}$ is the Roche-lobe radius and $B \simeq 10^4$. A fraction of the matter may be accreted by the secondary through Bondi-Hoyle accretion \citep{Bondi}. The system widens owing to loss of angular momentum and can remain too wide for Roche lobe overflow. In this case, mass transfer is not conservative and metals are released from the system. Because wind accretion only increases the secondary mass by a small amount and mass loss via winds increases the binary separation, this mechanism provides a possible formation scenerio for carbon-enhanced metal-poor stars with masses around $1\,M_\odot$ and no observed binary companion. These may be long period binaries. Wind accretion can be significant if the binary is close enough that the primary radius is close to the Roche Lobe radius during the early AGB phase. 

Using the enhanced mass-loss rate and Bondi-Hoyle accretion \citep{Bondi}, we estimate the amount of material that can be accreted from a $7\,M_\odot$ star. The mass accreted by the secondary \citep*{Hurley}
\begin{equation}
\dot{M_{2}}=- [\frac {GM_{2}} {v^{2}_{\rm w}}] \frac {\alpha} {2a^{2}} \frac {1}{(1+v^{2})^{3/2}} \dot{M_{1}},
\end{equation}
where $\dot{M_{1}}$ is the rate of mass change of primary, $M_{2}$ is mass of secondary and $a$ is the separation of two stars. The parameter $\alpha$ is taken to be 3/2 appropriate for Bondi-Hoyle accretion. The wind velocity of mass lost from primary is $v_{\rm{w}}$. This is difficult to determine, so it is simply taken to be the escape velocity from the surface, most likely an overestimate. The ratio between orbital velocity and wind velocity is $v$.  
 
Using the above, we calculate that a primodial $0.6\,M_\odot$ star accretes  $0.2\,M_\odot$ from the wind of a $7\,M_\odot$ star and the whole envelope of the primary is lost before thermal pulses begin. More detailed modelling and calculation of the secondary surface abundances are not carried out here. The abundances of the models are estimated for two extreme cases a) the accreted material mixes with the whole star and b) the accreted material remains in a surface layer. 

The abundances of C and N fit fairly well with the original data of \shortcite{Christlieb}. The abundances of C, N and O from the 3-D atmospheric analysis published by \shortcite*{collet} are significantly lower (by 0.8 dex or more) than the 1-D analysis (Table~7). Thus, it is possible that the observed high carbon and oxygen abundances of HE 0107-5240 are due to accretion from the wind of an AGB star. Compared with oxygen abundances given by \shortcite{Bessell}, oxygen is underproduced in the model. We investigate whether changes to the stellar physics could result in a better fit to the oxygen. First we use  the ${}^{12}\rm{C}(\alpha,\gamma){}^{16}\rm{O}$ reaction rate of \shortcite{Angulo} and find that increasing this rate by a factor of five is sufficient to raise the mass of ${}^{16}\rm{O}$ dredged up to the observed value at negligible expense of the ${}^{12}\rm{C}$. However, the upper limit of the rate given by \shortcite{Angulo} is only 50 per cent greater than the adopted one in the relevant temperature range. Though some still doubt this result \citep{Imbriani}, there is as yet no evidence for an unknown resonance in the reaction, so the current reaction rate is too low to reproduce the observation. Secondly we investigate the addition of convective overshooting as described in \shortcite{Polsov} but find the effect small because of the stability of the Schwarzschild boundary at the base of the convective envelope during dredge up. Other possibilities include a higher oxygen abundance in low-metallicity stars compared to solar. This would imply the star has an initial composition enriched by the earlier generation of stars which is possible given $[\rm{Fe}/\rm{H}] \approx -5$

Another notable difference is that our models have a surface enhancement of sodium by a factor of two which is due to the high intershell temperature of a $7\,M_\odot$ star. However sodium is observed to be enhanced by more than a factor of six. This can be fixed by either increasing the reaction rate of ${}^{22}\rm{Ne}(p,\gamma)$ which has an uncertainty of 10 or by lowering the reaction rate of ${}^{23}\rm{Na}(p,\alpha)$. Also, during thermal pulses, sodium is produced when ${}^{22}\rm{Ne}$ from the intershell reacts with protons in the hydrogen shell. Sodium could then be created at the base of the convective envelope by hot-bottom burning and would be brought to the surface in a $7\,M_\odot$ star. The surface carbon to nitrogen ratio may also change through hot-bottom burning but it would still be within the observational range after few thermal pulses. The strength of thermal pulses is too weak for carbon or $s$-process elements to be dredged up. Alternatively, the over-abundances of sodium may again be due to enrichement by an earlier generation of stars. 

The second star, HE 1327-2326, has strong enrichment in the neutron-capture element strontium but no barium. Strontium is usually  the product of the weak \textit{s}-process, which produces essentially no barium. This would fit the binary scenario if the mass accretion continues briefly after the start of some third dredge-up, when the surface of the primary is enriched in \textit{s}-process neutron-capture elements. However, the neutron flux has to be low, and the efficiency of the weak \textit{s}-process at low metallicity is uncertain \citep{Aoki}. Also our $Z=0$ models do not show third dredge up. Another possible explanation is that HE 1327-2326 was formed from a cloud enriched by a nearby supernovae which produced strontium via the \textit{r}-process \citep{Truran}. HE 1327-2326 is also much more  enriched in nitrogen which can be produced by hot-bottom burning in intermediate-mass AGB stars. On the whole, the abundances seem to suggest that binary mass transfer at a later AGB phase might fit the observed abundances of HE 0107-5240. Such stars, which require the solution of a number of complex numerical problems, will be modelled in future work.

\begin{table}
\begin{center}
\begin{tabular}[t] {c c c c c c}
\hline
Element & 1D log $\epsilon(X)$ & 3D log $\epsilon(X)$ & With mixing & Without mixing \\
\hline
C & 6.81 &  5.72  &   5.44   &   6.08\\
N & 4.93 &  2.91  &   3.10   &   3.71\\
O & 5.66 &  4.95  &   3.49   &   3.71\\
\hline
\end{tabular}

\caption{Observed abundances of HE 0107-5240 compared with binary models, log $\epsilon(X)= \rm log_{10}(N_{X}/N_{H})+12.0$}
\end{center}
\end{table}

\section{Conclusions}
Mass transfer to a companion of low mass during AGB phases results in chemically peculiar stars such as carbon-enhanced metal-poor stars. Mass accreted by the secondary can significantly enhance the surface abundances of the CNO isotopes and possibly neutron-capture elements. For example, the abundances of HE 0107-5240 can be modelled by wind accretion from a $7\,M_\odot$ primary during early AGB phases. HE 1327-2326 could also be formed in similar way but with mass transfer occuring at a later stage. If such mass transfer is non-conservative metals are also released to the interstellar medium after second dredge up and enrich the interstellar medium. This is possible even if the binary is sufficiently close that Roche lobe overflow is followed by common envelope evolution in which the envelope is lost from the system. These binary systems contribute more to the interstellar medium than a single star at the same stage of evolution because the envelope is lost earlier. However, the total contribution from a primary star in binary system is smaller than in a single system when the single star ends its life as a type 1.5 supernova. This would release a large yield of carbon, oxygen, nickel and iron. The combined contribution of binary stars and single stars depends, not only on the binary fraction and separation in the early Universe, but also the mass-loss rate and evolutionary endpoints of single stars of zero metallicity. The more stars in close binaries, the smaller would be the total nitrogen yield because the hot bottom burning in the AGB phase would be curtailed.

\section{Acknowledgements}
HBL thanks PPARC for his Dorothy Hodgkin Scholarship. RJS and CAT thank Churchill College for their fellowships. We would also like to thank the anonymous referee for useful comments and suggestions.

\bibliography{HBL}

\label{lastpage}

\end{document}